\documentclass[floats,aps,twocolumn,showpacs]{revtex4}
\usepackage{dcolumn}
\usepackage{epsfig}
\newcommand {\la} {\langle}
\newcommand {\ra} {\rangle}
\newcommand {\beq} {\begin{eqnarray}}
\newcommand {\eeqn} [1] {\label{#1} \end{eqnarray}}%

\begin{document}
%
%\initfloatingfigs
%

%\tighten
\title{On the existence of a bound tetraneutron}


\author{
N.\ K.\ Timofeyuk
}                     % Do not remove

\address{
Physics Department, University of Surrey, Guildford,
Surrey GU2 7XH, England, UK
}

\date{Received: \today}

\begin{abstract}
Following   recent work in which events which may correspond 
to a bound tetraneutron  ($^4$n) were  observed, 
it is pointed out  that from the theoretical perspective
the two-body nucleon-nucleon force
cannot  by itself bind  four neutrons, even if it can bind a dineutron.
 A very strong phenomenological four-nucleon (4N)
force is needed in order to bind the tetraneutron. Such a 4N force,  
if it existed, would   bind   $^4$He by about 100 MeV.
Alternative   experiments  such as  ($^8$He,$^4$n)  
are proposed to search for the tetraneutron.

\end{abstract}

%}
%
\maketitle

 In a recently reported experiment \cite{marques,Orr} events were observed 
that exhibit the characteristics of a multineutron cluster liberated in the 
breakup of $^{14}$Be, most probably in the channel $^{10}$Be+$^4$n.
The lifetime of order 100~ns or longer suggested by this measurement, would 
indicate that the tetraneutron is particle stable.  If confirmed,
 the existence 
of a bound tetraneutron is of great importance as it would challenge our 
understanding of nuclear few-body systems and nucleon-nucleon 
(NN) interactions.
Here I would like to make several observations in this context.
 
{\bf I}

As discussed in Refs \cite{marques,Orr} the breakup $^{14}$Be $\rightarrow$ 
$^{10}$Be+$^4$n represents an attractive channel to search for
 a tetraneutron.
The nucleus $^{10}$Be is strongly bound and the four neutron separation 
energy 
for $^{14}$Be is only  about 5 MeV  with
respect to the  $^{14}$Be $ \rightarrow \,^{10}$Be + 4n breakup.
It is known that in  an external potential well neutrons
can form drops \cite{ndrops}, \cite{ndrops1}.
In the attractive field of $^{10}$Be the last four neutrons could form
a tetraneutron which might be liberated in the breakup.

 In earlier  experiments, the tetraneutron was searched
using  heavy-ion transfer reactions such as 
 $^7$Li($^{11}B,^{14}O$)$^4$n \cite{Bel},
$^7$Li($^7Li,^{10}$C)$^4$n \cite{Al} and
double exchange  reaction $^4$He($\pi^-,\pi^+$)$^4$n  \cite{U}, \cite{G}.
These reactions should be strongly suppressed by
spectroscopic reasons and a
negative outcome from them could be easily anticipated.

{\bf II}

No proper {\em ab-initio} four-body calculations of the tetraneutron
with
realistic two-body and three-body NN forces are known to the author. However,
several   theoretical calculations of the tetraneutron are available.\\
$(i)$ In Ref. \cite{bev} the tetraneutron,
studied  in the translationally invariant, symmetrized
oscillator basis   using the $4\hbar\omega$ model space,
was found to be unbound by 18.5 MeV. However, the oscillator
basis is not appropriate for the description of unbound systems.\\
$(ii)$
No bound tetraneutron was found in Ref. \cite{gorb}  within the
angular potential functions method with semirealistic NN interactions.
No search for the four-body resonance state was carried out there.\\
$(iii)$ 
No bound tetraneutron was found in Ref. \cite{varga} within the
stochastic variational method on a correlated Gaussian basis
for a range of simple effective NN
potentials. No search for the four-body resonance state was
carried out there.\\
$(iv)$ A search of the four-body resonance in the lowest order of the
hyperspherical functions method (HSFM) gave a null result \cite{BBKE},
\cite{SRV}.\\
$(v)$ The energy behaviour of the
eigen phases, studied in Ref. \cite{GNO} within
the HSFM using  the $K_{max}$ = 6 model space, led the authors to 
the conclusion
that the tetraneutron may exist as a resonance in the four-body
continuum at an energy of about $1 - 3$ MeV.
However, such a conclusion is not convincing because no convergency of
the eigen phases with an
increase of the hyperangular momentum has
been achieved in these calculations.
Besides, a clear indication of the resonance has been seen with only
one of the NN potentials used in the calculations,
namely with the Volkov effective NN force V1 \cite{volkov}.
Volkov effective NN
interactions reproduce the experimental binding energy of another
four-body system, $^4$He. However, their singlet even and triplet even
components  are equal, and therefore with these potentials
 a singlet dineutron has exactly the same binding energy as the deuteron.
In the particular case of V1, the dineutron is bound by 0.547 MeV.
Therefore, V1 cannot be used in  calculations of the multineutron systems.
Another NN potential, used in Ref. \cite{GNO}, namely that
of Reichstein and Tang  (RT) \cite{RT}, reproduces the n-p triplet
and p-p singlet scattering  lengths and does not bind a dineutron. 
With this potential the energy derivatives of the eigen phases
 monotonically decrease with increasing energy.
This means that
resonances in the four neutron system are absent, at least within the
model space considered.

To understand whether the RT potential can produce any resonance
or bound state
if the model space of the HSFM is increased,
I have calculated   the hyperradial potentials for the tetraneutron
up to $K_{max}$ = 16 using the technique developed in Ref. \cite{T}.
  The RT potential has only central even components
$V_{10}$ and $V_{01}$ and
the odd potentials are usually obtained from them
using  an arbitrary parameter $u$ so that
$V_{00} =  (u-1) V_{01}$ and $V_{11} =(u-1)V_{10}$.
The present calculations have been performed
with $u = 1$ which corresponds to  no-interaction in the
odd partial waves. The diagonalised hyperradial potentials $V_{diag}(\rho)$,
calculated for different model spaces $K_{max}$,
are plotted in Fig.1 as a function of hyperradius $\rho$.
Such potentials can be used to obtain  a quick approximate
solution for the binding energy in the extreme adiabatic approximation
of the HSFM \cite{fabre}.
One can see that the
$V_{diag}(\rho)$ have almost converged. They are purely repulsive,
monotonically decreasing with the hyperradius,
and do not show any sign of local attractive pockets.
Therefore,   the RT potential can neither
bind the tetraneutron   nor produce any resonances.
To get a bound tetraneutron, the value of $u$ should be increased up to
$u=2.3$. However, with this value, the isotopes $^4$H and $^5$He must
become bound with respect to the t + n and $^4$He + n thresholds
by about 6.5 MeV and 32 MeV respectively.
In reality, these nuclei are unbound by  3 and 0.9 MeV.

\begin{figure}[h]
\centerline{
        \psfig{figure=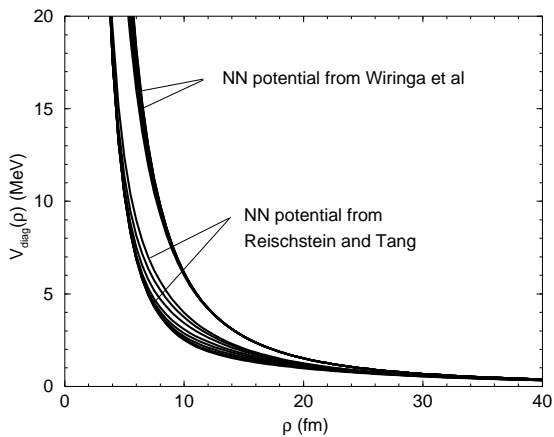,width=0.47\textwidth}
        }

\caption{  Diagonalised hyperradial potentials $V_{diag}(\rho)$
of the tetraneutron
 calculated with $K_{max}$  = 2, 4, ..., 16
 with two different central NN potentials.}
\end{figure}
%---------------------------------------------------------------------------

For comparison, the   calculations of the hyperradial potentials
have been performed with Volkov NN potential V1  as well.
The diagonalised hyperradial potentials $V_{diag}(\rho)$,
 shown  in Fig. 2,  reveal
local    attractive pockets  for $K_{max} > 6$. These pockets
become negative for $K _{max} > 12$, but they are too  shallow
to form a  bound state with respect to the four-body decay.
Although  with $K_{max}$ = 16 convergence has not yet been reached, the 
general
trend seen in Fig. 2, suggests that it is unlikely, that
a tetraneutron, bound with respect to the $^2$n + $^2$n decay,
may exist. To get a bound tetraneutron,
the Majorana parameter $m$ should be changed from its standard value of 0.6
to $m = -0.2$. 
Such a change  provides E($^4$n) = $-$1.2 MeV and
does not influence the binding energy of $^4$He, however, it
binds $^4$H and $^5$He with respect to the t + n  
and $^4$He + n thresholds by  7 MeV 25 MeV respectively.

\begin{figure}[h]
\centerline{
        \psfig{figure=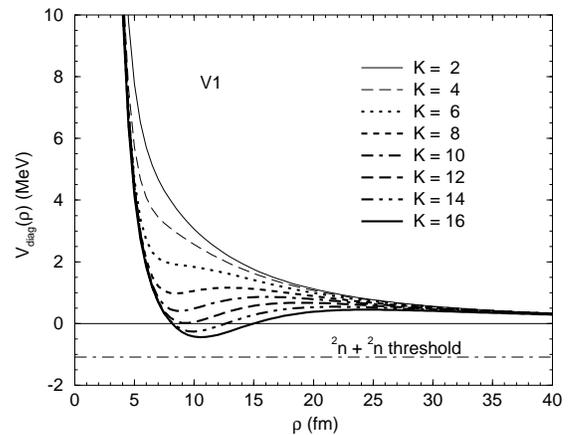,width=0.47\textwidth}
        }

\caption{  The same as in Fig.1 for the Volkov potential V1. }
\end{figure}
%-------------------------------------------------------------------------
%

And finally,   calculations with  the realistic
two-body isospin-conserving central NN interaction Argonne $v_4$
\cite{wiringa} have been performed as well
and  the corresponding hyperradial potentials $V_{diag}(\rho)$
are shown in Fig.1. These purely repulsive potentials have almost 
converged and
their behaviour excludes any possibility of either a bound
or a resonance state.

{\bf III}

The results of the theoretical calculations suggest that  the tetraneutron
must be unbound and most likely should not exist as a resonance,
at least when only two-body central forces are considered.
It is remarkable that even when an effective NN potential
binds the dineutron (the case of V1), it still cannot bind
two  dineutrons, although a resonance in such a system should exist.
In order for the two-body central force  to be able to bind the tetraneutron,
a huge unphysical attraction  should be introduced in  the triplet
odd potential. Such an attraction would strongly overbind $^4$H, $^5$He
and other known $A > 4$ nuclei. 
The   reason for necessity of huge attraction lies in the relative
number of nucleon pairs in even and odd states allowed by the Pauli 
principle.
For the tetraneutron
with  $L = 0$ and $S = 0$ the probabilities $P$ to find a pair
of nucleon in the singlet even (01) and triplet
odd (11) states are equal, $P_{01}=P_{11}$ = $\frac{1}{2}$. 
 For the tetraneutron with $L = 1$ and $S = 1$,  
these probabilities are $P_{01} = \frac{1}{3}$ and $P_{11} = \frac{2}{3}$.
The NN force  is
not sufficiently attractive in the
singlet even state,  and   is repulsive in the
triplet odd state, which leads to the absence of the
tetraneutron.
For comparison, for the $L = 0$, $S = 0$ state of the closed shell
nucleus $^4$He $P_{10} = P_{01} = \frac{1}{2}$  and for the
$L = 1$, $S = 1$ state of $^4$H
$P_{10} = P_{01} = P_{11}=\frac{1}{3}$. 
The binding energies of $^4$He and $^4$H
are  $-$28.3 MeV and $-$5.2 MeV respectively.
It is clear that  the binding of a nucleus is strongly
correlated with the probabilities to find a pair of nucleons in even states,
especially in the triplet even state (10) where the $n-p$ bound state exists.

 {\bf IV}

The fact that, despite   the theoretical predictions made with central
two-body forces, six   events
 $^{14}$Be $\rightarrow \, ^{10}$Be + $^4$n  have possibly been observed,
means  that   either
 non-central interactions   and/or three-nucleon (3N) forces,
and/or   a strong four-nucleon (4N)  force may  bind the
tetraneutron. If not, then a different interpretation of the six
events   from Refs \cite{marques,Orr} should be looked for.

The first possibility should be rejected, because the  calculations of Ref.
\cite{gorb} indeed included the spin-orbit and tensor forces.
The second possibility does not look convincing either because
the contribution of the 3N potential to the
 binding  energies of the neutron drops
$^7$n and $^8$n, calculated with  Green's function Monte
Carlo Methods in Ref. \cite{pieper},
 is    about 1 to 5 MeV which
is small    compared to $^4$He and to
other $A = 7$ and $A=8$ nuclei. A similar  contribution from the 3N
force would be too  small to bind the tetraneutron.
 Therefore, if what has been seen in the $^{14}$Be experiment
was indeed the tetraneutron, it could indicate at   presence of a
4N force.

%---------------------------------------------------------------------------

\begin{figure}[t]
\centerline{
        \psfig{figure=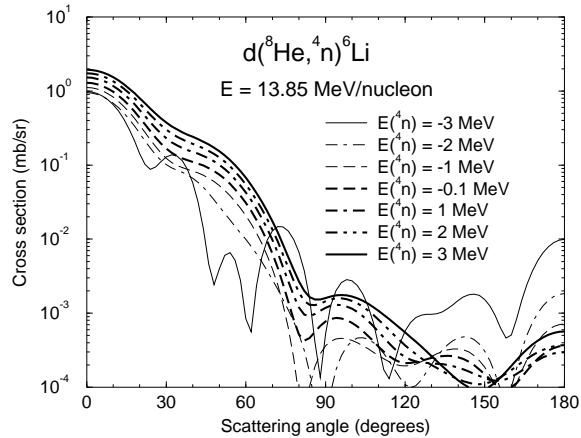,width=0.47\textwidth}
        }

\caption{ The DWBA $\alpha$-transfer cross sections of the 
d($^8He,^4n$)$^6$Li
reaction as a function of the tetraneutron binding energy.}
\end{figure}
%-------------------------------------------------------------------------

At present, the results of the {\em ab-initio} calculations of
the  $3 \leq A \leq 8$ nuclei suggest that a  4N force
is not needed to fit  the observed binding energies of light nuclei
at the 1$\%$ level \cite{pieper}. Therefore, either the 4N
contributions are smaller than 1$\%$ for these nuclei or parts of the
3N forces are mocking up their effects \cite{pieper}. On the other hand,
the latest models of the
3N force  used to calculated the binding energies
of these nuclei
have not yet been tested in the description of  polarization observables
of the low-energy $Nd$ scattering where the contribution of the
3N force is very important \cite{kiev}, \cite{cadman}.
Besides, the latest numerically accurate calculations employing a 3N
force do not reproduce  the proton
analyzing power for the p-$^3$He scattering \cite{viviani}.
The  simultaneous fit of the $Nd$ and $p-^3He$ polarization
observables and  of binding energies of the lightest nuclei
may leave  room for a 4N force.

In order to get an idea of what   strength
is needed for the 4N force,
the HSFM calculations  of this paper have been repeated
with the RT potential using a  4N force   simulated
by the  potential
$V_{4N}(\rho) = W_0 e^{-\alpha \rho}$. The values  $\alpha = 0.7, 1.2$
and 1.5 $fm^{-1}$,
used in calculations, were the same as the range of the phenomenological
term of the  3N spin-orbit force   introduced in Ref. \cite{kiev}.
The tetraneutron becomes bound if
 the corresponding values $W_0$ are equal to
$-410$, $-1460$ and $-2530$  MeV respectively. These values are two
orders of magnitude larger as
compared to the values $-1$, $-10$, and $-20$ MeV obtained
for  the 3N spin-orbit force \cite{kiev}.
For the Volkov potential V1, these strengths are similar:
$W_0$ = $-320$,  $-1565$ and $-2900$ MeV.
If the same 4N force existed in $^4$He,
the binding energy of $^4$He, calculated with V1,
 would be $-88$, $-82$ and $-134$ MeV respectively.
Therefore, the 4N force, if it exists, should be strongly
$T$-dependent. In principle, such a strong $T=2$ 4N force
could be noticable in $A-Z\ge 4$ nuclei, especially near the neutron
edge of stability, where the number of the $T=2$ four-nucleon states
 increases.
However,  
HSFM calculations of the neutron-rich
helium isotopes $^{6,8}$He, performed  with   V1 \cite{T},
do not indicate a need for a strong 4N force.

To get a real understanding of the role of the  
4N force, {\em ab-initio} calculations of the tetraneutron
with realistic 2N and 3N forces must be performed.
The HSFM is the best method
for such  calculations because
 it is  well suited to systems without bound subsystems, which
 is the case for the tetraneutron, and because it
 can also  treat a four-body continuum.

%---------------------------------------------------------------------------

\begin{figure}[t]
\centerline{
        \psfig{figure=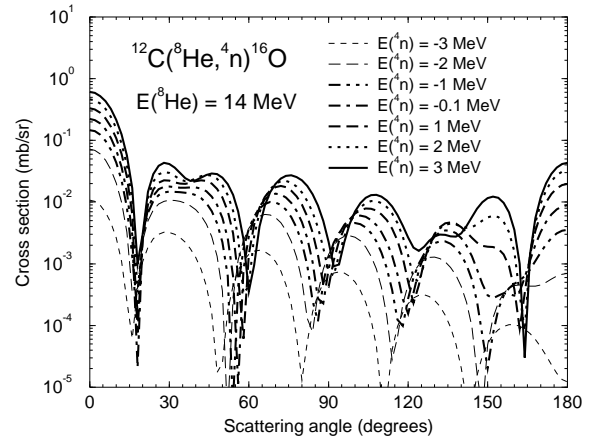,width=0.47\textwidth}
        }

\caption{  The same as Fig. 3, but for the
$^{12}$C($^8He,^4n$)$^{16}$O reaction.}
\end{figure}
%-------------------------------------------------------------------------

{\bf V}

Since the discovery of the tetraneutron would  most certainly require a
revision of  modern realistic models of the three-nucleon interaction,
it is extremely important to confirm or reject the results reported in
Ref. \cite{marques}. Therefore, independent searches for the tetraneutron
in alternative reactions should continue.
Among such reactions,   $\alpha$ transfer
reactions $A(^8He,^4n)A+\alpha$ deserve  special attention.
Using a light target and detecting a product nucleus $A + \alpha$, one can
study not only a hypothetical tetraneutron bound state but also the
four-neutron  continuum as well.

The spectroscopic factor $S$ for the $\la^8He|^4He\otimes^4n\ra$ overlap
can be estimated using the translation invariant shell model \cite{TS}.
If the   probability amplitudes of the $|L=0,S=0 \ra$
and $|L=1,S=1 \ra$ configurations are $\alpha_0$ and $\alpha_1$
 in $^8$He and $\beta_0$ and $\beta_1$ in $^4$n respectively,
then      $S = \left(\sqrt{\frac
{2}{3}} \alpha_0\beta_0 + \sqrt{\frac{5}{6}} \alpha_1\beta_1 \right)^2$.

The choice of the target $A$ must be determined by kinematic and
spectroscopic considerations. It should be light enough for  
residual nucleus
$A+\alpha$ to come out of the target and its wave function should
overlap strongly
with the $A+\alpha$ wave function. The obvious targets of this type
are the deuteron, triton, $^3$He and $^4$He.
 The list of other light targets
with large spectroscopic factors $S_{\alpha}$ is given in Ref. \cite{kurath}.

As an example, in Figs. 3 and 4 the DWBA $\alpha$-transfer
cross sections of the
 d($^8He,^4n$)$^6$Li and $^{12}$C($^8He,^4n$)$^{16}$O reactions
are shown as a function of the four-body binding energy of the
hypothetical tetraneutron. The energies of these reactions have been
chosen based on availability of optical potentials in the entrance
and exit channels, which have been taken from \cite{perey} and
\cite{bec}. In addition, the  energy choice for  the $^{12}$C 
target was also motivated by
experimental measurements of the similar reaction 
$^{12}$C($^8Li,\alpha$)$^{16}$N which revealed very large cross sections 
at small angles, about 25 mb/sr \cite{bec}. 
 The latest
theoretical value $S_{\alpha}$ = 0.87 \cite{NWS} has been used
 for  $^6$Li and $S_{\alpha}$ for $^8$He was taken to be $\frac{2}{3}$.  
According to Figs. 3 and 4,  
the  d($^8He,^4n$)$^6$Li  reaction is better suited for the tetraneutron
search because  its cross sections are not strongly influenced by
unknown tetraneutron binding energy.
The cross sections
and $^8$He beam intensities make this experiment feasible in the very near
future. 
In contrast, the cross sections of the 
$^{12}$C($^8He,^4n$)$^{16}$O reaction decrease dramatically with
increasing the binding of the tetraneutron and become small  
if the tetraneutron is bound. \\

Summarizing, due to the large probability for a pair of neutrons
to be in the triplet odd state,
the two-body nucleon-nucleon force
cannot  by itself bind  four neutrons, even if it can bind a dineutron.
 A very strong phenomenological $T$-dependent four-nucleon  
force is needed in order to bind the tetraneutron.
Alternative   experiments  such as  ($^8$He,$^4$n)
are proposed to search for the tetraneutron.

\vskip 0.5 cm
I am grateful to W. Catford for providing me with the text of Ref. 
\cite{marques} prior to the publication. I am also grateful to N.Orr
and W. Catford
for useful comments concerning my paper.
I also thank K. Varga for providing me with the gaussian expansion
of the Argonne NN potential.

\end{document}